\documentclass[reqno,11pt]{amsart}
\usepackage{graphicx}
\usepackage{slashed}
\usepackage{amscd}
\usepackage{tensor}
\usepackage{amssymb}
\usepackage{esint}
\usepackage{enumitem}
\usepackage{accents}
\usepackage[mathscr]{eucal}
\numberwithin{equation}{section}
\allowdisplaybreaks[1]

\title[Self-Adjointness of a Non-Uniformly Elliptic Dirac Hamiltonian]{Self-Adjointness of the Dirac Hamiltonian \\
for a Class of Non-Uniformly Elliptic \\ Boundary Value Problems}

\author[F.\ Finster]{Felix Finster}
\author[C.\ R\"oken]{Christian R\"oken \\ \\ December 2015}
\thanks{Supported by the DFG research grant ``Dirac Waves in the Kerr Geometry: Integral Representations, Mass Oscillation Property and the Hawking Effect.''}
\address{Fakult\"at f\"ur Mathematik \\ Universit\"at Regensburg \\ D-93040 Regensburg \\ Germany}
\email{finster@ur.de, Christian.Roeken@mathematik.ur.de}

\newtheorem{Def}{Definition}[section]
\newtheorem{Thm}[Def]{Theorem}
\newtheorem{Prp}[Def]{Proposition}
\newtheorem{Lemma}[Def]{Lemma}

\newtheorem{Corollary}[Def]{Corollary}
\newtheorem{Example}[Def]{Example}

\newcommand{\Thanks}{\vspace*{.5em} \noindent \thanks}
\newcommand{\beq}{\begin{equation}}
\newcommand{\eeq}{\end{equation}}
\newcommand{\Proof}{\begin{proof}}
\newcommand{\QED}{\end{proof} \noindent}
\newcommand{\QEDrem}{\ \hfill $\Diamond$}

\newcommand{\la}{\langle}
\newcommand{\ra}{\rangle}

\newcommand{\Sl}{\mbox{$\prec \!\!$ \nolinebreak}}
\newcommand{\Sr}{\mbox{\nolinebreak $\succ$}}

\newcommand{\C}{\mathbb{C}}
\newcommand{\R}{\mathbb{R}}
\newcommand{\1}{\mbox{\rm 1 \hspace{-1.05 em} 1}}

\newcommand{\N}{\mathbb{N}}

\newcommand\B{{\mathscr{B}}}
\newcommand{\U}{\text{\rm{U}}}

\newcommand{\Dir}{{\mathcal{D}}}
\DeclareMathOperator{\supp}{supp}

\newcommand{\D}{\mathscr{D}}

\newcommand{\scrM}{\mycal M}
\newcommand{\scrN}{\mycal N}

\DeclareFontFamily{OT1}{rsfso}{}
\DeclareFontShape{OT1}{rsfso}{m}{n}{ <-7> rsfso5 <7-10> rsfso7 <10-> rsfso10}{}
\DeclareMathAlphabet{\mycal}{OT1}{rsfso}{m}{n}

\begin{document}

\maketitle

\begin{abstract}
We consider a boundary value problem for the Dirac equation
in a smooth, asymptotically flat Lorentzian manifold
admitting a Killing field which is timelike near and tangential to the boundary.
A self-adjoint extension of the Dirac Hamiltonian is constructed.
Our results also apply to the situation that the space-time includes horizons,
where the Hamiltonian fails to be elliptic.
\end{abstract}

\tableofcontents

\section{Introduction}
Let~$(\scrM, g)$ be a smooth, oriented and time-orientated Lorentzian 
spin manifold of dimension~$d \geq 3$ with boundary~$\partial \scrM$.
Moreover, we make the following assumptions:
\begin{itemize}
\item[(i)] The manifold~$(\scrM, g)$ is {\em{asymptotically flat}} with one asymptotic end.
\item[(ii)] There is a {\em{Killing field}}~$K$ which is {\em{tangential to
and timelike on~$\partial \scrM$}}.
\item[(iii)] The integral curves of~$K$, defined by the differential equation
\[ \dot{\gamma}(t) = K\big( \gamma(t) \big) \:, \]
exist for all~$t \in \R$.
\item[(iv)] There exists a {\em{spacelike hypersurface}}~$\scrN$ 
with compact boundary~$\partial \scrN$ with the property
that every integral curve~$\gamma$ in~(iii) intersects~$\scrN$ exactly once.
\end{itemize}
These assumptions imply that~$\scrM$ and its boundary~$\partial \scrM$ have the product structures
\beq \label{splitting}
\scrM = \R \times \scrN \qquad \text{and} \qquad \partial \scrM = \R \times \partial \scrN \:.
\eeq
Note that our assumptions also imply that the metric~$g$ is smooth up to the boundary~$\partial \scrM$,
thus inducing on~$\partial \scrN$ a $(d-2)$-dimensional Riemannian metric.
%

For clarity, we now mention two well-known special cases.
If~$\partial {\scrN}$ is empty and~${\scrN}$ is complete, the product
structure~\eqref{splitting} implies that~$(\scrM, g)$ is {\em{globally hyperbolic}}.
Moreover, if~$K$ is timelike in the asymptotic end, the manifold is {\em{stationary}}.
However, we point out that we merely assume that the Killing field~$K$ is timelike on the boundary~$\partial \scrM$,
but it does not need to be timelike everywhere.

In order to get a better geometric understanding of the above setting,
we now construct a convenient coordinate system.
Choosing the parametrization of each curve~$\gamma$ such that~$\gamma(0) \in \scrN$,
we obtain a global coordinate function~$T$ defined by
\beq \label{Tdef}
T \::\: \scrM \rightarrow \R \qquad \text{with} \qquad T\big(\gamma(t)\big) = t\:.
\eeq
The level sets of this time function give rise to a foliation~$\scrN_t := T^{-1}(t)$ by spacelike hypersurfaces
with~$\scrN_0 =\scrN$.
Moreover, the integral curves give rise to isometries
\[ \Phi_t \::\: \scrN \rightarrow \scrN_t \:, \qquad \Phi_t\big(\gamma(0) \big) = \gamma(t) \:. \]
Choosing coordinates~$x$ on~$\scrN$, the mapping~$x \circ \Phi_t^{-1}$ gives
coordinates on~$\scrN_t$.
Complementing this coordinate system by the function~$t=T$,
we obtain coordinates~$(t, x)$
with~$t \in \R$ and~$x \in \scrN$ such that~$K=\partial_t$.
In these coordinates, the line element takes the form
\[ ds^2 = g_{ij} \:dx^i\, dx^j
= a(x)\, dt^2 + b_\alpha(x)\, dt\, dx^\alpha - \big(g_\scrN(x)\big)_{\alpha \beta}\: dx^\alpha\, dx^\beta \:, \]
where~$a$ and~$b_\alpha$ are smooth functions, and~$g_\scrN$ is the induced Riemannian metric on~$\scrN$. Here we denote the space-time indices by latin letters~$i,j \in \{0, 1, 2, \ldots, d-1 \}$, whereas
spatial indices are denoted by greek letters~$\alpha, \beta \in \{1,2, \ldots, d-1 \}$.
This coordinate system can be understood as describing an observer who is
co-moving along the flow lines of the Killing field.
We note that  in the regions where~$K$ is timelike, the function~$a(x)$ is positive,
and the metric is stationary. This is the case if~$x$ is near the boundary~$\partial \scrN$.
However, away from~$\partial \scrN$, the function~$a(x)$ could be negative, in which case the
metric is {\em{not stationary}}, and~$t$ is {\em{not}} a {\em{time coordinate}}. This situation is illustrated in the
following example.

\begin{Example} \label{exkerr} {\bf{(Kerr geometry in Eddington-Finkelstein-type coordinates)}} {\em{
In the recent paper~\cite{eddington}, horizon-penetrating Eddington-Finkelstein-type coordinates
\[ (\tau, r, \theta, \phi) \qquad \text{with} \qquad \tau \in \R, \;r \in \R^+, \;\theta \in (0,\pi), \;\phi \in (0, 2 \pi) \]
are introduced in the four-dimensional
non-extreme Kerr geometry. In these coordinates, the line element takes the form
\begin{align*}
ds^2 & = \biggl(1 - \frac{2 M r}{\Sigma} \biggr) d\tau^2 - \frac{4 M r}{\Sigma} \bigl(dr - a \sin^2 \theta\: d\phi\bigr) d\tau  \\
& - \biggl(1 + \frac{2 M r}{\Sigma}\biggr) \bigl(dr - a \sin^2 \theta \:d\phi\bigr)^2
- \Sigma \, d\theta^2 - \Sigma \sin^2 \theta\: d\phi^2 \:,
\end{align*}
where~$\Sigma = r^2 + a^2 \cos^2 \theta$. Moreover, $M$ and~$aM$ denote the mass and the
angular momentum of the black hole, respectively. The surfaces~$r_\pm=M\pm \sqrt{M^2-a^2}$
are the {\em{event horizon}} and the {\em{Cauchy horizon}} of the black hole. Note that these
coordinates are regular on and across the horizons. The two Killing fields describing the stationarity and
axisymmetry of the Kerr geometry are~$\partial_\tau$ and~$\partial_\phi$.

We choose a radius~$r_0<r_-$ inside the Cauchy horizon and let
\[ \scrM = \{ r > r_0 \} \:, \qquad \scrN = \{ \tau=0, \;r > r_0 \}
\qquad \text{and} \qquad \partial \scrM = \{ r = r_0 \} \:. \]
Direct computation shows 
that the Killing field~$\partial_\tau$ is {\em{not}} everywhere timelike on~$\partial \scrM$
(due to an ergo-like region inside the Cauchy horizon; for details see~\cite{hamilton}).
But taking~$K$ as a suitable linear combination of~$\partial_\tau$ and~$\partial_\phi$,
\[ K = \partial_\tau + b \,\partial_\phi \]
with a real constant~$b=b(r_0) \neq 0$, it turns out that~$K$ is a Killing field which satisfies all the above assumptions.
This Killing field is spacelike near spatial infinity. }} \QEDrem
\end{Example}

We next formulate the Dirac equation. To this end, we choose an arbitrary spin structure
and let~$S\scrM$ be the corresponding {\em{spinor bundle}}.
It is a vector bundle with fibers~$S_p \scrM \simeq \C^f$, $f \in \scrM$,
where the dimension~$f$ is given by~$f = 2^{[d/2]}$
(where~$[.]$ is the lower Gauss bracket; thus~$f=4$ in dimensions~$d=4$ or~$5$).
Each fiber is endowed with an indefinite inner product of signature~$(f/2,f/2)$, referred to as
{\em{spin scalar product}} and denoted by
\[ \Sl .|. \Sr_p \::\: S_p \scrM \times S_p \scrM \rightarrow \C \:. \]
The geometric Dirac operator~$\Dir$ takes the form
\beq \label{Dirgeom}
\Dir = i \gamma^j \nabla_j \:,
\eeq
where the Dirac matrices~$\gamma^j$ are related to the metric by
the anti-commutation relations
\[ \big\{ \gamma^j, \gamma^k \big\} = 2\,g^{jk}\:\1_{S_p\scrM}\:, \]
and~$\nabla$ is the metric connection on the spinor bundle
(for more details see~\cite{lawson+michelsohn}).
In order to allow for an external potential (like for example an electromagnetic potential),
instead of~\eqref{Dirgeom} we shall consider the more general Dirac operator
\beq \label{Dir}
\Dir = i \gamma^j \nabla_j + \B \:,
\eeq
where~$\B$ is a smooth matrix-valued potential which we assume to be symmetric
with respect to the spin scalar product, i.e.\ $\Sl \phi | \B \psi \Sr = \Sl \B \phi | \psi \Sr$.

We are interested in solutions~$\psi$ of the Dirac equation of mass~$m$
\beq \label{Deq}
(\Dir - m) \psi = 0\:,
\eeq
with the Dirac operator according to~\eqref{Dir}. In order to analyze the
dynamics of Dirac waves, it is useful to write the Dirac equation in the
Hamiltonian form
\beq \label{Hamilton}
i \partial_t \psi = H \psi \:,
\eeq
where~$H$ is the Dirac Hamiltonian given by
\beq \label{DirH}
H = -\big(\gamma^t \big)^{-1} \big( i \gamma^\alpha \nabla_\alpha + \B - m \big) \:.
\eeq
Taking the domain of definition
\[ \D(H) = C^\infty_0 \big( \accentset{\circ}{\scrN}, S\scrM \big) \: ,\]
where~$\accentset{\circ}{\scrN} = \scrN\setminus\partial \scrN$ denotes the interior of~$\scrN$,
this Hamiltonian is indeed symmetric (i.e.\ formally self-adjoint) with respect to the scalar product
\[ (\psi | \phi)_\scrN = \int_\scrN \Sl \psi | \slashed{\nu} \phi \Sr_x \: d\mu_\scrN(x)\:, \]
where~$\nu$ is the future-directed normal on~$\scrN$ and~$d\mu_\scrN$ is the volume form
on~$(\scrN, g_\scrN)$.
This can be verified with the following computation. Using current conservation together
with the fact that the metric coefficients do not depend on the coordinate~$t$,
we obtain
\begin{align*}
0 &= \partial_t \big(\psi(t) \,\big|\, \phi(t) \big)_\scrN = (\dot{\psi} | \phi)_\scrN + (\psi | \dot{\phi})_\scrN \\
&= \big((-i H \psi) \,\big|\, \phi \big)_\scrN + \big( \psi \big| (-i H \phi) \big)_\scrN
= i \big( (H \psi | \phi)_\scrN - (\psi | H \phi)_\scrN \big)\:.
\end{align*}

In order to pose the Cauchy problem for the Dirac equation~\eqref{Deq}, one needs
to specify initial and boundary conditions.
We choose initial data which is smooth and compactly supported,
\beq \label{init}
\psi|_{\scrN} = \psi_0 \in C^\infty_0(\scrN, S\scrM) \:.
\eeq
Moreover, we impose the boundary conditions
\beq \label{boundary}
(\slashed{n} - i)\: \psi|_{\partial \scrM} = 0 \:,
\eeq
where the slash denotes Clifford multiplication, and~$n$ is the inner normal on~$\partial \scrM$ (meaning
that for every~$p \in \partial \scrM$
there is a curve~$c : [0,\delta) \rightarrow \scrM$ with~$c(0)=p$ and~$\dot{c}(0)=n(p)$).
Clearly, the initial data must be compatible with the boundary
conditions, meaning that~$(\slashed{n} - i)\: \psi_0|_{\partial \scrM} = 0$.
These boundary conditions, which are very similar to those introduced in~\cite[Section~2]{tkerr},
have the effect that Dirac waves are reflected on~$\partial \scrM$.
They can also be understood in analogy to the chiral boundary conditions
in~\cite{dougan+mason, gibbons+hawking+horowitz}.
The difference is that, instead of the intrinsic Dirac operator on
the hypersurface, we here consider the Hamiltonian obtained from the Dirac
operator in space-time by separating the $t$-dependence.
This gives rise to the additional factor~$(\gamma^t)^{-1}$ in~\eqref{DirH}.
Our boundary conditions~\eqref{boundary} can be understood as
an adaptation of the chiral boundary conditions in~\cite{dougan+mason, gibbons+hawking+horowitz}
to the Hamiltonian~\eqref{DirH}.

The boundary conditions~\eqref{boundary} must be incorporated in the functional analytic setting.
To this end, one extends the domain of definition to
\beq \label{domright}
\D(H) = \big\{ \psi \in C^\infty_0(\scrN, S\scrM) \quad \text{with} \quad (\slashed{n} - i)\: \psi|_{\partial \scrN} = 0 
\big\} \:.
\eeq
Then the operator~$H$ is again symmetric, as the following consideration shows. First, we rewrite the scalar product $(\psi | H \phi)_\scrN$ in a more convenient form. Applying the relations~$(\gamma^t)^2 = g^{t t} \1_{S_x\scrM}$
and~$\slashed{\nu} = \gamma^t/\sqrt{g^{t t}}$, we obtain
\[ \slashed{\nu} (\gamma^t)^{- 1} = \frac{1}{g^{t t}} \, \slashed{\nu} \, \gamma^t = \frac{\1_{S_x\scrM}}{\sqrt{g^{t t}}} \:. \]
Making use of the form of the Hamiltonian (\ref{DirH}), this leads to
\[ (\psi | H \phi)_\scrN = - i \int_\scrN \Sl \psi | \gamma^{\alpha}  \nabla_{\alpha} \phi \Sr_x \:
\frac{1}{\sqrt{g^{t t}}}\;d\mu_\scrN(x) + (\text{lower order terms}) \, .\]
Now a direct computation of the boundary terms yields
\[ (\psi | H \phi)_\scrN - (H \psi | \phi)_\scrN = i \int_{\partial \scrN} \Sl \psi | \slashed{n} \phi \Sr_x
\: \frac{1}{\sqrt{g^{t t}}}\; d\mu_{\partial \scrN}(x) \]
(we note that the angular derivatives do not give rise to boundary terms because~$\partial \scrN$ is compact without
boundary).
Using the boundary conditions in~\eqref{domright}, for all~$x \in \partial \scrN$ we obtain
\[ i \,\Sl \psi | \phi \Sr_x = \Sl \psi | \slashed{n} \phi \Sr_x = \Sl \slashed{n} \psi | \phi \Sr_x =
-i \,\Sl \psi | \phi \Sr_x \:, \]
proving that the boundary values indeed vanish. This shows that~$H$ is symmetric.

In order to solve the Cauchy problem and to analyze the long-time behavior of its solutions,
it is of central importance to construct a self-adjoint extension of~$H$.
Namely, with such a
self-adjoint extension at hand, the solution of the Cauchy problem for the Dirac equation~\eqref{Deq}
with initial values~\eqref{init} and boundary conditions~\eqref{boundary} can be expressed using the
spectral theorem for self-adjoint operators as
\[ \psi(t) = e^{-i t H}\: \psi_0 = \int_{\sigma(H)} e^{-i \omega t} \: dE_\omega\, \psi_0 \:. \]
This formula is also the starting point for a detailed analysis of the long-time behavior of~$\psi$
using spectral methods, similar as carried out in the exterior region of Kerr geometry in~\cite{tkerr, decay}.
In the present paper, we succeed in constructing a self-adjoint extension:
\begin{Thm} \label{thmmain}
The Dirac Hamiltonian~\eqref{DirH} with domain of definition
\beq \label{domright2}
\D(H) = \Big\{ \psi \in C^\infty_0(\scrN, S\scrM) \quad \text{with} \quad (\slashed{n} - i) \big(H^p \psi\big) \big|_{\partial \scrN} = 0 \quad \text{for all~$p \in \N_0$} \Big\}
\eeq
is essentially self-adjoint.
\end{Thm} \noindent
Note that the domain~\eqref{domright2} is smaller than~\eqref{domright}. This is preferable because
we want that the Cauchy problem has a global solution in~$C^\infty_0(\scrN, S\scrM)$
(see the strategy of our proof as described at the end of this section).

We conclude this section by putting our result into the context of previous work,
and explaining the strategy of our proof. The Cauchy problem and the problem of constructing
a self-adjoint extension of the Dirac Hamiltonian have been studied in several simpler situations:
\begin{itemize}[leftmargin=2em]
\item[(i)] If~$\scrN$ is a complete manifold without boundary, the Cauchy problem
\label{i}
can be solved using the theory of symmetric hyperbolic systems (see for example~\cite{john, taylor3}).
In this construction, one works with local charts with local time functions.
Since the resulting local solutions coincide in the regions where the charts overlap, this procedure
gives rise to a unique, global smooth solution in~$\scrM$.
Then, restricting this solution to the spacelike hypersurfaces of constant~$t$,
one obtains a family of time evolution operators
\[ \U_{t',t} \::\: C^\infty \big( \{t\} \times \scrN, S\scrM \big) \rightarrow C^\infty
\big( \{t'\} \times \scrN, S\scrM \big) \:, \]
which form a group. This makes it possible to apply~\cite{chernoff73} to conclude
that the Hamiltonian is essentially self-adjoint on~$C^\infty(\scrN, S\scrM)$.
\item[(ii)] In the ultrastatic situation
\[ ds^2 = dt^2 - \big(g_\scrN\big)_{\alpha \beta}\: dx^\alpha\, dx^\beta \:, \]
the Hamiltonian can be written as
\[ H = \begin{pmatrix} 0 & \Dir_\scrN \\ \Dir_\scrN & 0 \end{pmatrix} \:, \]
where~$\Dir_\scrN$ is the intrinsic Dirac operator on~$\scrN$.
This makes it possible to apply the results in the Riemannian setting
as worked out in detail in~\cite{bartnik+chrusciel}.
\item[(iii)] In the static situation
\[ ds^2 = a(x)\: dt^2 - \big(g_\scrN\big)_{\alpha \beta}\: dx^\alpha\, dx^\beta \:, \]
the Hamiltonian can be written as
\[ H = \sqrt{a(x)} \begin{pmatrix} 0 & \Dir_\scrN \\ \Dir_\scrN & 0 \end{pmatrix} + \text{(zero-order terms)}
\:.\]
Introducing a suitable scalar product on the Dirac wave functions,
this Hamiltonian is again symmetric, making it possible to again apply the
results of~\cite{bartnik+chrusciel}.
\end{itemize}
In the situation under consideration here, there is the
major complication that the Hamiltonian is in general not uniformly elliptic,
so that the methods in~\cite{bartnik+chrusciel} no longer apply.
In order to explain the problem, we now consider the principal symbol
of the Dirac Hamiltonian. According to~\eqref{DirH}, the principal symbol
takes the form
\[ P(x,\xi) = -i \big(\gamma^t \big)^{-1} \gamma^\alpha\, \xi_\alpha \:. \]
The ellipticity condition states that the principal symbol should be bounded from below by
\[ \big\|P(x, \xi) \big\| \geq \delta \,\|\xi\|^2 \]
for a suitable constant~$\delta>0$ (for basics on the principal symbol and the connection
to ellipticity see for example~\cite[Section~5.11]{taylor1}). In order to verify whether this condition holds,
it is most convenient to compute the determinant of the principal symbol. Namely,
\[ \det P(x,\xi) = \det\big((\gamma^t )^{-1}\big) \: \det \big(\gamma^\alpha\, \xi_\alpha \big) \:, \]
and using that
\[ (\gamma^t )^{-1} (\gamma^t )^{-1} = \frac{\1_{S_x \scrM}}{g^{tt}} \:,\qquad
\gamma^\alpha\, \xi_\alpha\; \gamma^\beta\, \xi_\beta = g^{\alpha \beta}\, \xi_\alpha \,\xi_\beta\,\1_{S_x \scrM} \:, \]
we obtain
\[ \det P(x,\xi) = \bigg( \frac{g^{\alpha \beta}\, \xi_\alpha \,\xi_\beta}{g^{tt}} \bigg)^{f/2} \:. \]
This computation shows that the Hamiltonian fails to be elliptic if~$g^{\alpha \beta}\, \xi_\alpha \,\xi_\beta=0$
for a non-zero~$\xi$.
In the example of the Kerr metric in Eddington-Finkelstein-type coordinates~\cite{eddington}, this is the
case precisely on the event and Cauchy horizons.
More generally, the points where the Hamiltonian fails to be elliptic can be used as the definition
of the {\em{horizons}} of our space-time. Thus we face the major problem that the
{\em{Hamiltonian is not elliptic on the horizons}}.

Our strategy to solve this problem is to split up the solution of the Cauchy problem into two separate problems:
Near the boundary, we rewrite the problem in a form where the results in~\cite{bartnik+chrusciel} apply.
Away from the boundary, however, we use the theory of symmetric hyperbolic systems.
Making essential use of finite propagation speed, adding the two solutions gives rise to
a unique solution of our boundary value problem for small times. By iterating the procedure, we
get unique global, smooth solutions, making it possible to proceed as in~(i) above
by applying~\cite{chernoff73}.

\section{A Double Boundary Value Problem}
As a technical tool for the proof of Theorem~\ref{thmmain}, we need to show that the
Cauchy problem~\eqref{Deq}, \eqref{init} with boundary conditions~\eqref{boundary} has global smooth solutions.
Our method is to split up the Cauchy problem into two separate problems near
and away from the boundary. In preparation, we now introduce additional boundary conditions
on a suitable surface~$Y$ near~$\partial \scrM$.
We work in Gaussian normal coordinates
in a tubular neighborhood in~$\scrN$ of~$\partial \scrN$.
Thus for any~$p \in \partial \scrN$, we let~$c_p(r)$ for~$0 \leq r < r_{\max}(p)$
be the geodesic in~$\scrN$ with the initial conditions
\[ c_p(0) =p \qquad \text{and} \qquad c_p'(0) = u \:, \]
where~$u \in T_p\scrN$ is the inner normal to~$\partial \scrN$.
Since~$\partial \scrN$ is compact, we can choose~$r_{\max}>0$ independent of~$p$ to obtain
a mapping
\[ c \::\: [0, r_{\max}) \times \partial \scrN \rightarrow \scrN \:,\qquad c(r,p) = c_p(r) \:. \]
Applying the implicit function theorem, possibly by decreasing~$r_{\max}$, we can arrange
that~$c$ is a diffeomorphism. We introduce the sets obtained from~$\partial \scrN$ by the
geodesic flow by
\[ \partial \scrN(r) = c(r,\partial \scrN) \:. \]
Choosing coordinates~$\Omega=(\vartheta_1, \ldots, \vartheta_{d-2})$ on~$\partial \scrN$
gives a corresponding coordinate system~$(r, \Omega)$ on~$\scrN$.
In these coordinates, the metric on~$\scrN$ takes the form
\beq \label{gNform}
(g_\scrN)_{\alpha \beta} = \begin{pmatrix} 1 & 0 \\ 0 & g_{\partial \scrN(r)} \end{pmatrix}
\qquad \text{and thus} \qquad
(g_\scrN)^{\alpha \beta} = \begin{pmatrix} 1 & 0 \\ 0 & \big( g_{\partial \scrN(r)} \big)^{-1} \end{pmatrix} \:.
\eeq
Taking again~$t$ as the time coordinate, we obtain a coordinate system~$(t,r,\Omega)$
with~$t \in \R$, $r \in [0, r_{\max})$ of~$\scrM$ which describes a neighborhood of~$\partial \scrM$.

We now introduce the following new boundary value problem. Let~$X$ be the space-time region
\[ X = \big\{ (t,r,\Omega) \:\big|\: 0 \leq r \leq r_{\max}/2 \big\} \:. \]
This is a Lorentzian manifold whose boundary~$\partial X$ consists of~$\partial \scrM$ as well as the
$(d-1)$-dimensional surface
\[ Y := \big\{ (t, r_{\max}/2, \Omega) \big\} \:. \]
Possibly by decreasing~$r_{\max}$, we can arrange that~$K$ is timelike in~$X$,
implying that~$Y$ is a timelike surface. The inner normal on~$Y$ is again denoted by~$n$.
We consider the initial value problem
\beq \label{initbox}
(\Dir - m) \,\psi = 0 \quad \text{in~$X$}\:,\qquad \psi|_{\scrN} = \psi_0 \in C^\infty(\scrN \cap X, S\scrM)\:,
\eeq
with the boundary conditions
\beq \label{boundbox}
(\slashed{n} - i)\,\psi|_{\partial X} = 0  \: ,
\eeq
where~$\partial X = \partial \scrM \cup Y$ now has two components.
It is again useful to rewrite the Dirac equation in the Hamiltonian form~\eqref{Hamilton}
with the Hamiltonian~\eqref{DirH}.
In order to take into account the boundary conditions, we now choose the domain of
definition as the Sobolev space
\beq \label{DH}
\D(H) = \big\{ \psi \in W^{1,2}(X \cap \scrN, S\scrM) \;\big|\; 
(\slashed{n} - i)\: \psi|_{\partial X \cap \scrN} = 0 \big\} \:.
\eeq
The next proposition gives a spectral decomposition of~$H$.

\begin{Prp} \label{prpdoublebasis}
There is a countable orthonormal basis~$(\psi_n)_{n \in \N}$, $\psi_n \in \D(H)$, of eigenfunctions of~$H$.
\end{Prp}
\Proof Our method is to apply the abstract spectral theorem given in~\cite[Theorem~4.1]{bartnik+chrusciel}.
The task is to verify the spectral conditions~($\mathcal{C}$0)--($\mathcal{C}$4), which in our
setting are stated as follows:
\begin{itemize}[leftmargin=2.5em]
\item[($\mathcal{C}$0)] $H \::\: \D(H) \rightarrow L^2(X)$ is linear and bounded in the~$W^{1,2}$-topology
on~$\D(H)$.
\item[($\mathcal{C}$1)] The G\r{a}rding inequality holds: There exists a constant~$C$ such
that for all~$\psi \in \D(H)$,
\beq \label{Garding}
\|\psi\|_{W^{1,2}(X \cap \scrN)}^2 \leq C \int_{X \cap \scrN} \left( \Sl H \psi | \slashed{\nu} H \psi \Sr_x
+ \Sl \psi | \slashed{\nu} \psi \Sr_x \right) d\mu_{\scrN}(x) \:.
\eeq
\item[($\mathcal{C}$2)] Weak solutions are strong solutions (``elliptic regularity''): If~$\phi \in L^2(X \cap \scrN)$
satisfies
\[ \int_{X \cap \scrN} \Sl H \psi | \slashed{\nu} \phi \Sr_x\: d\mu_\scrN(x)
= 0 \qquad \text{for all~$\psi \in \D(H)$} \:, \]
then~$\phi \in \D(H)$.
\item[($\mathcal{C}$3)] $H$ is symmetric, i.e.\ for all~$\psi, \phi \in \D(H)$,
\[ \int_{X \cap \scrN} \Sl \psi | \slashed{\nu} H \phi \Sr_x\:d\mu_{\scrN}(x)
= \int_{X \cap \scrN} \Sl H \psi | \slashed{\nu} \phi \Sr_x\:d\mu_{\scrN}(x) \:. \]
\item[($\mathcal{C}$4)] $\D(H)$ is dense in~$L^2(X \cap \scrN)$.
\end{itemize}
The validity of condition~($\mathcal{C}$0) follows immediately from the fact
that~$H$ is a differential operator of first order.
The symmetry property~($\mathcal{C}$3) was verified after~\eqref{domright}.
The denseness property~($\mathcal{C}$4) is obvious.
In order to verify the G\r{a}rding inequality~($\mathcal{C}$1), we
exploit the specific form of the Hamiltonian~\eqref{DirH}.
The contribution to~$\Sl H \psi | \slashed{\nu} H \psi \Sr_x$ involving first derivatives squared is estimated by
\[ \Sl (\gamma^t)^{-1} \gamma^\alpha \nabla_\alpha \psi
\,|\, \slashed{\nu} \:(\gamma^t)^{-1} \gamma^\beta \nabla_\beta \psi \Sr_x 
\geq c\: g_\scrN^{\alpha \beta}\: \|\nabla_\alpha \psi\| \:  \|\nabla_\beta \psi\| \]
(for a suitable constant~$c>0$), where we used that the inner product~$\Sl .| \slashed{\nu} . \Sr$ is positive definite
and that the matrices~$(\gamma^t)^{-1}$ and~$\gamma^\alpha$ are uniformly bounded on~$X$.
Introducing the notation~$\|\nabla \psi\|^2 = g_\scrN^{\alpha \beta}\: \|\nabla_\alpha \psi\| \:  \|\nabla_\beta \psi\|$
and estimating the coefficients of the lower order terms by suitable constants~$d_1, d_2>0$, we obtain the estimate
\begin{align*}
\Sl H \psi | \slashed{\nu} H \psi \Sr_x &\geq c\: \|\nabla \psi\|^2
- d_1 \, \|\nabla \psi\|\: \|\psi\| - d_2\, \|\psi\|^2 \\
&\geq \frac{c}{2}\: \|\nabla \psi\|^2 - \left(\frac{d_1^2}{2c}+d_2 \right) \|\psi\|^2 \:.
\end{align*}
Hence
\begin{align*}
\|\psi\|_{W^{1,2}(X \cap \scrN)}^2 &= \int_{X \cap \scrN} \Big( \|\nabla \psi\|^2 + \|\psi\|^2 \Big)\: d\mu_{\scrN}(x) \\
&\leq \int_{X \cap \scrN} \left( \frac{2}{c}\:\Sl H \psi | \slashed{\nu} H \psi \Sr_x
+ \frac{2}{c}\bigg( \frac{d_1^2}{2c}+d_2 \bigg) \:\|\psi\|^2 + \|\psi\|^2 \right) d\mu_{\scrN}(x) \:.
\end{align*}
This shows that condition~(${\mathcal{C}}1)$ holds.

It remains to derive the regularity condition~($\mathcal{C}$2).
This consists of two parts: the interior regularity and the regularity at the boundary.
For the interior regularity, we need to show that the operator~$H$ is uniformly
elliptic (see~\cite[Theorem~3.7]{bartnik+chrusciel}).
To this end, we make use of the fact that the Killing field~$K$ is timelike in~$X$.
As a consequence, we can use it to define a norm on the spinors by
\[ \|\psi(x) \|_x^2 := \Sl \psi | \gamma^t \slashed{K} \gamma^t \psi \Sr_x \:. \]
Using this norm, we have
\begin{align*}
\big\| \xi_\alpha\, \big((\gamma^t)^{-1} \gamma^\alpha\big) \psi)(x) \big\|_x^2
&= \xi_\alpha \,\xi_\beta\; \Sl (\gamma^t)^{-1} \gamma^\alpha \psi \,|\, \gamma^t \slashed{K} \gamma^t\,
(\gamma^t)^{-1} \gamma^\beta \psi \Sr_x \\
&= \xi_\alpha \xi_\beta\; \Sl \gamma^\alpha \psi \,|\, \slashed{K} \gamma^\beta \psi \Sr_x \:.
\end{align*}
Since~$\slashed{K}=\gamma_t$, this matrix anti-commutes with the matrices~$\gamma^\beta$.
Therefore,
\begin{align*}
\big\| \xi_\alpha\, \big((\gamma^t)^{-1} \gamma^\alpha\big) \psi)(x) \big\|_x^2
&=-\xi_\alpha \,\xi_\beta\; \Sl \gamma^\beta \gamma^\alpha \psi \,|\, \slashed{K} \psi \Sr_x \\
&= -g^{\alpha \beta}\: \xi_\alpha\, \xi_\beta\; \Sl \psi \,|\, \slashed{K} \psi \Sr_x
= -g^{\alpha \beta}\: \xi_\alpha\, \xi_\beta\; \big\|(\gamma^t)^{-1} \psi \big\|_x^2 \:,
\end{align*}
showing explicitly that~$H$ is uniformly elliptic. To see that the matrix~$(\gamma^t)^{-1}$
is uniformly bounded, we note that, using Cramer's rule,
\[ \big( (\gamma^t)^{-1} \big)^2 = \frac{\1_{S_x \scrM}}{g^{tt}} = -\frac{\det g_{\alpha \beta}}{g_{tt}}
= -\frac{\det g_{\alpha \beta}}{\la K, K \ra} \:, \]
which is indeed bounded because~$K$ is timelike in~$X$.

The remaining proof of the boundary regularity is a subtle point, which we now treat in detail.
We first note that, by localizing with a test function and using the interior regularity, it suffices
to consider weak solutions whose support is in a small neighborhood of~$\partial \scrN$
or~$Y \cap \scrN$. Since both cases can be treated in the same way, we may assume that
the solution vanishes identically outside a small neighborhood of~$\partial \scrN$.
Our goal is to apply~\cite[Theorem~5.11]{bartnik+chrusciel}. Simplifying the statement of this
theorem and adapting it to our setting, this theorem gives boundary regularity for
boundary value problems of the form
\begin{align}
Lu &= f \label{Luf} \\
Pu|_{\partial \scrN} &= 0 \:.
\end{align}
Here~$P$ is a projection operator on~$L^2(\partial \scrN)$.
Moreover, $L$ is the differential operator
\[ L = \partial_r + A + B \:, \]
where the operators~$A$ and~$B$ are of the following form.
The operator~$A : W^{1,2}(\partial \scrN) \rightarrow L^2(\partial \scrN)$ is an angular differential 
operator which is independent of~$r$ and satisfies again the above spectral conditions~($\mathcal{C}$0)--($\mathcal{C}$4).
The operator~$B : W^{1,2}(X \cap \scrN) \rightarrow L^2(X \cap \scrN)$, on the other hand,
should be such that its first-order terms vanish on~$\partial \scrN$.

The first step is to rewrite the Dirac equation and the boundary conditions in the required form.
Suppose that~$\psi$ is a weak solution of the inhomogeneous equation~$H \psi = f$ satisfying the boundary
conditions~\eqref{boundbox}. Moreover, assume that~$\psi$ and~$f$ are supported in a small neighborhood
of~$\partial \scrN$. In order to implement the boundary conditions on~$\partial \scrN$, we choose
the projection operator~$P$ as
\[ P = \frac{1}{2} \big(i \slashed{n} + 1 \big) \:.\]
Next, using~\eqref{DirH}, one can write the differential equation~\eqref{Luf} as
\[ \Big( \partial_r + (\gamma^r)^{-1} \big(
\gamma^{\vartheta_1} \partial_{\vartheta_1} + \cdots + \gamma^{\vartheta_{d-2}} \partial_{\vartheta_{d-2}}
\big) + E \Big) \psi =
i (\gamma^r)^{-1} \gamma^t f  \:, \]
where~$(\vartheta_a)_{a=1,\ldots, d-2}$ are again coordinates on~$\partial \scrN$,
and~$E$ is a zero-order operator. We choose
\begin{align}
A &= (\gamma^r)^{-1} \big(\gamma^{\vartheta_a} \partial_{\vartheta_a} \big) \big|_{\partial \scrN} + Z \label{Adef} \\
B &= (\gamma^r)^{-1} \big(\gamma^{\vartheta_a} \partial_{\vartheta_a}  \big) + E - A \:,
\end{align}
where~$Z$ is a zero-order operator on~$\partial \scrN$ to be determined below.

The crucial point is to show that by a suitable choice of the scalar product
and the zero-order operator~$Z$, we can arrange that
the operator~$A$ is symmetric. We choose the scalar product as
\beq \label{Ksprod}
\la .|. \ra_{\partial \scrN} = \int_{\scrN} \Sl \,.\, | \slashed{K} \,.\, \Sr_x\: d\mu_{\partial \scrN}
\eeq
(since~$K$ is timelike near~$\partial \scrN$, this inner product is indeed positive definite).
Using the form of the metric~\eqref{gNform} in our Gaussian normal coordinate system,
the following anti-commutation relations hold,
\[ \{ \slashed{K}, \gamma^r \} = 2 \, g\indices{_t^r} = 2\,\delta_t^r = 0 \:,\qquad
\{ \slashed{K}, \gamma^{\vartheta_a} \} = 0 \:,\qquad
\{ \gamma^r, \gamma^{\vartheta_a} \} = 0\:. \]
As a consequence, the matrices~$(\gamma^r)^{-1} \gamma^{\vartheta_a}$
are anti-symmetric with respect to the scalar product~\eqref{Ksprod}.
Thus, setting
\[ Z = -\frac{1}{2} \:\big( A_0 - A_0^* \big) \qquad \text{with} \qquad
A_0 := (\gamma^r)^{-1} \big( \gamma^{\vartheta_a} \partial_{\vartheta_a} \big)
\big|_{\partial \scrN} \:, \]
where the star denotes the formal adjoint with respect to the scalar product~\eqref{Ksprod},
the operator~$Z$ is indeed a multiplication operator. Moreover, using the above formulas for~$Z$
and~$A_0$ in~\eqref{Adef}, one sees that~$A=(A_0+A_0^*)/2$, which is obviously symmetric.
Finally, it is clear by construction
that the restriction of~$B$ to~$\partial \scrN$ is a multiplication operator.

From this construction, it is obvious that the operator~$A$ has the above properties~($\mathcal{C}$0),
($\mathcal{C}$3) and~($\mathcal{C}$4). In order to prove the G\r{a}rding inequality~($\mathcal{C}$1)
and the elliptic regularity~($\mathcal{C}$2), we make use of the anti-commutation relations
\[ \Big\{(\gamma^r)^{-1} \,\gamma^{\vartheta_a}, (\gamma^r)^{-1} \,\gamma^{\vartheta_b} \Big\} =
- \frac{1}{g^{rr}} \:\big\{ \gamma^{\vartheta_a}, \gamma^{\vartheta_b} \big\} = -2\: \frac{g^{{\vartheta_a} {\vartheta_b}}}{g^{rr}} \:. \]
Hence the operator~$A^2$ is of the form
\[ A^2 = \frac{1}{g^{rr}}\: \Delta_{S^{d-2}} + (\text{lower order terms}) \:. \]
This is an elliptic operator on a bounded domain. Standard elliptic theory implies~($\mathcal{C}$1)
and~($\mathcal{C}$2).
\QED

The spectral decomposition of Proposition~\ref{prpdoublebasis} implies that the mixed
initial/boun\-da\-ry value problem~\eqref{initbox}, \eqref{boundbox} has
a unique weak solution in~$W^{1,2}(X \cap \scrN, S\scrM)$ given by
\beq \label{psiser}
\psi(t,x) = \sum_{n=1}^\infty c_n\, e^{-i \omega_n t}\: \psi_n(x) \qquad \text{with} \qquad
c_n = \int_{X \cap \scrN} \Sl \psi_n | \slashed{\nu} \psi_0 \Sr_y\:d\mu_{\scrN}(y) \:,
\eeq
where~$\omega_n$ is the eigenvalue of~$\psi_n$.
In order to apply~\cite{chernoff73}, we want a solution which is smooth for all times.
We now state the corresponding necessary and sufficient conditions.
\begin{Lemma} \label{lemmasmooth}
Suppose that~$\psi_0$ satisfies the conditions
\beq \label{bp}
(\slashed{n} - i) \big(H^p \psi_0 \big) \big|_{\partial \scrN} = 0 \qquad \text{for all~$p \in \N_0\:.$} 
\eeq
Then the solution~$\psi$ of the mixed initial/boundary value problem~\eqref{initbox}, \eqref{boundbox}
is in the class~$C^\infty_\textnormal{sc}(\scrM, S\scrM)$, where the index ``sc'' denotes solutions of
space-like compact support (i.e.~$\supp \psi(t, .)$ is a compact subset of~$\scrN$
for all~$t \in \R$). Conversely, if a solution of the mixed initial/boundary value problem is smooth,
then~$\psi_0$ satisfies the conditions~\eqref{bp}.
\end{Lemma}
\Proof Let~$\psi$ be the solution of the mixed initial/boundary value problem~\eqref{initbox}, \eqref{boundbox}
for~$\psi_0$ satisfying~\eqref{bp}. In order to show that~$\psi$ is smooth, it clearly suffices
that all time derivatives of~$\psi$ exist and are smooth in~$x$. To this end, we consider the
partial sums of~\eqref{psiser}
\[ \psi^N(t,x) = \sum_{n=1}^N c_n\, e^{-i \omega_n t}\: \psi_n(x) \]
for given~$N \in \N$. Differentiating $p$ times with respect to~$t$ gives
\[ (i \partial_t)^p \psi^N(t,x) = \sum_{n=1}^N \omega_n^p c_n\, e^{-i \omega_n t}\: \psi_n(x) \:. \]
Furthermore,
\[ \omega_n^p c_n = \int_{X \cap \scrN} \Sl H^p \psi_n | \slashed{\nu} \psi_0 \Sr_y\:d\mu_{\scrN}(y)
= \int_{X \cap \scrN} \Sl \psi_n | \slashed{\nu} \big(H^p \psi_0 \big) \Sr_y\:d\mu_{\scrN}(y) \:, \]
where we iteratively integrated by parts and used the boundary conditions~\eqref{bp}.
Since the function~$\tilde{\psi}_0 := H^p \psi_0$ is again in~$\D(H)$ given by~\eqref{DH},
we can take the limit~$N \rightarrow \infty$ to conclude that
\[ (i \partial_t)^p \psi(t,x) = \sum_{n=1}^\infty \tilde{c}_n\, e^{-i \omega_n t}\: \psi_n(x) \qquad \text{with} \qquad
\tilde{c}_n = \int_{X \cap \scrN} \Sl \psi_n | \slashed{\nu} \tilde{\psi}_0 \Sr_y\:d\mu_{\scrN}(y) \:. \]
This shows that~$\psi$ is indeed a smooth solution.

Assume conversely that~$\psi$ is a smooth solution to the mixed initial/boundary value problem~\eqref{initbox}, \eqref{boundbox}. Then~$(\slashed{n}-i) \psi(t)|_{\partial \scrN}=0$ for all~$t$. Differentiating $p$ times
with respect to~$t$ gives
\[ 0 = (i \partial_t)^p \Big( (\slashed{n}-i) \psi(t)|_{\partial \scrN} \Big) \Big|_{t=0}
= (\slashed{n}-i) \big( H^p \psi_0 \big) \big|_{\partial \scrN} \:, \]
proving~\eqref{bp}.
\QED

\section{Solution of the Cauchy Problem}
We now return to the Cauchy problem~\eqref{Deq}, \eqref{init} with boundary conditions~\eqref{boundary}.
Thus we seek for solutions of the Dirac equation in the Hamiltonian form
\beq \label{Deqhamilton}
i \partial_t \psi = H \psi \quad \text{in $\scrM$} \:,
\eeq
with initial and boundary values
\beq \label{initboundary}
\psi|_\scrN = \psi_0 \qquad \text{and} \qquad (\slashed{n} - i)\: \psi|_{\partial \scrM} = 0 \:,
\eeq
where the initial data is in~$\D(H)$ as given in~\eqref{domright2}, i.e.
\beq \label{initclass}
\psi_0 \in  \Big\{ \psi \in C^\infty_0(\scrN, S\scrM) \quad \text{with} \quad 
(\slashed{n} - i) \big(H^p \psi\big) \big|_{\partial \scrN} = 0 \quad \text{for all~$p \in \N_0$} \Big\} \:.
\eeq

\begin{Lemma} \label{lemmalocal}
There is~$\varepsilon>0$ such that the mixed initial/boundary value problem~\eqref{initboundary},
\eqref{initclass} has a unique solution~$\psi$ in the class
\[ \big\{ \psi \in C^\infty_0([0, \varepsilon) \times \scrN, S\scrM) \quad \text{with} \quad (\slashed{n} - i)\:
(H^p \psi)|_{[0, \varepsilon) \times \partial \scrN} = 0   \quad \text{for all~$p \in \N_0$}
\big\} \:. \]
\end{Lemma}
\Proof Near~$\partial \scrN$, we again choose the Gaussian normal coordinate system where
the metric takes the form~\eqref{gNform}. Moreover, we choose~$\varepsilon$ so small that
the future development~$J^\vee$ of initial data sets has the properties
\begin{align}
J^\vee \Big( \big\{(0,r,\Omega) \:\big|\: r< r_{\max}/4 \big\} \Big) \cap \big( \{\varepsilon\} \times \scrN \big) &\subset
\big\{(\varepsilon,r,\Omega) \:\big|\: r< r_{\max}/2 \big\} \label{small1} \\
J^\vee \Big( \big\{(0,r,\Omega) \:\big|\: r > r_{\max}/8 \big\} \Big) \cap \big( \{\varepsilon\} \times \scrN \big) &\subset
\big\{(\varepsilon,r,\Omega) \:\big|\: r>0 \big\} \:. \label{small2}
\end{align}

\begin{figure}
\includegraphics[width=8.6cm]{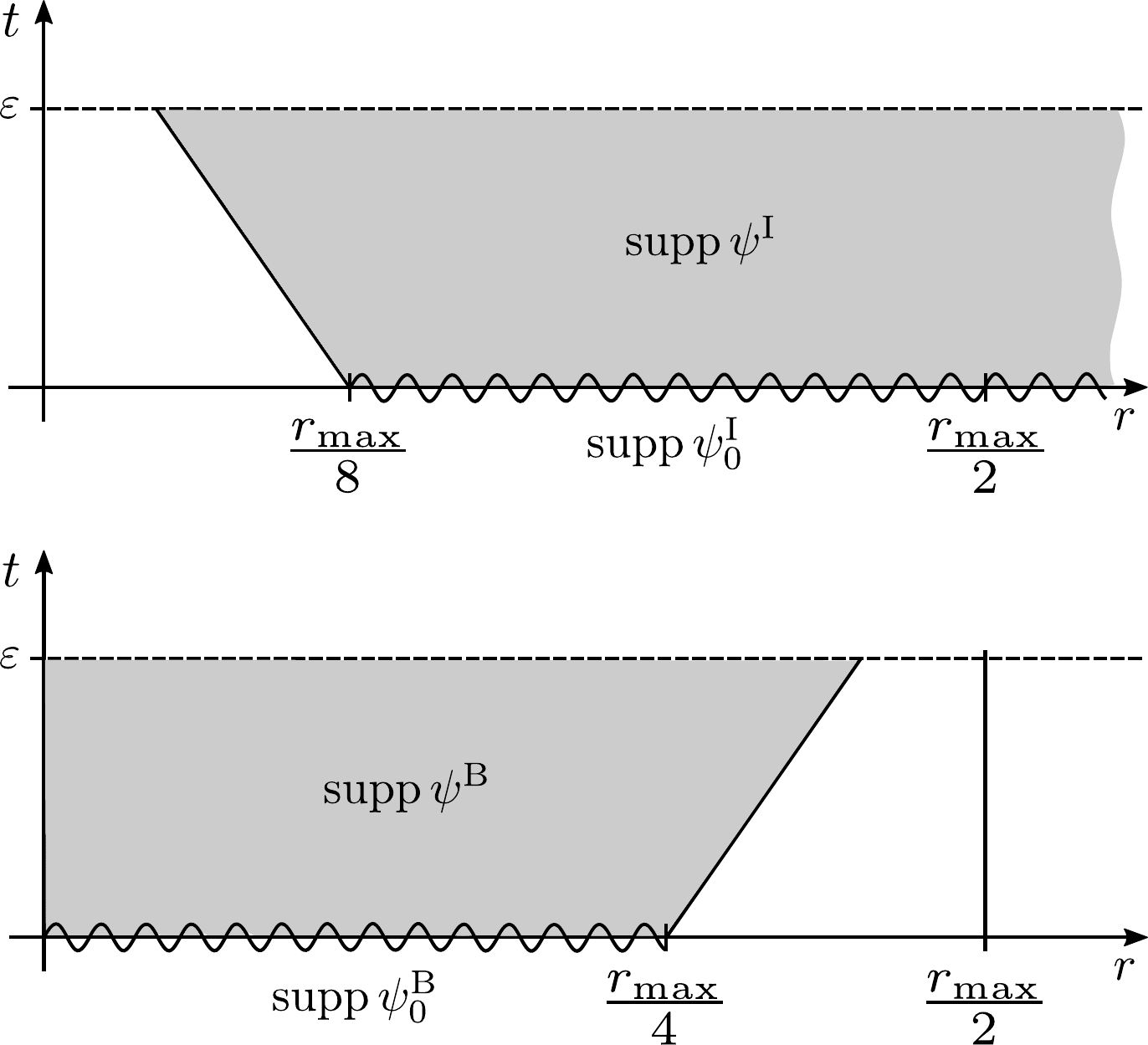}
\caption{Decomposition of the solution of the Cauchy problem.} \label{fig}
\end{figure}

We next decompose the initial data into a contribution~$\psi_0^\text{B}$ near the boundary~$\partial \scrN$
and a contribution~$\psi_0^\text{I}$ supported in the interior of~$\scrN$,
\[ \psi_0 = \psi_0^{\text{B}} + \psi_0^{\text{I}} \:. \]
To this end, we let~$\eta \in C^\infty_0\big( (-r_{\max}/4, r_{\max}/4) \big)$ 
be a test function with~$\eta|_{[0, r_{\max}/8]} \equiv 1$ and set (see Figure~\ref{fig})
\[ \psi_0^\text{B} := \eta(r)\, \psi_0 \qquad \text{and} \qquad \psi_0^\text{I} := \psi_0 - \psi_0^\text{B} \:. \]

We take~$\psi_0^\text{I}$ as initial value problem for the Dirac equation without boundary conditions,
\[ i \partial_t \psi^\text{I} = H \psi^\text{I} \;\;\;\;\text{in $\accentset{\circ}{\scrM}$} 
\:,\qquad\quad  \psi^\text{I}|_\scrN = \psi^\text{I}_0 \:. \]
Using the theory of symmetric hyperbolic equations (see~\cite[Section~5.3]{john}, 
\cite[Section~16]{taylor3}, \cite[Section~7]{ringstroem} or~\cite[Chapter~5]{intro}),
this initial value problem has a unique solution~$\psi^\text{I}$
in the class~$C^\infty_\textnormal{sc} \big([0, \varepsilon) \times \scrN)$
(just as explained in~(i) on page~\pageref{i}).
Note that, due to finite propagation speed and~\eqref{small2}, the solution vanishes
identically near~$\partial \scrM$ (see the top picture in Figure~\ref{fig}).

Next we take~$\psi_0^\text{B}$ as initial values for the double boundary value problem, i.e.
\beq \label{mixedbox}
i \partial_t \psi^\text{B} = H \psi^\text{B} \;\;\;\text{in $X$} \:,\qquad \psi^\text{B}|_\scrN = \psi^\text{B}_0
\:,\qquad (\slashed{n} - i)\: \psi^\text{B}|_{\partial \scrM \cup Y} = 0 \:.
\eeq
According to Lemma~\ref{lemmasmooth}, this mixed initial/boundary value problem has
a smooth solution which satisfies the initial and boundary conditions in~\eqref{mixedbox} pointwise
(this solution even satisfies the stronger boundary conditions in~\eqref{bp}).
Moreover, due to finite propagation speed and~\eqref{small1}, we know that the solution~$\psi^\text{B}$ vanishes
near the boundary~$\{r=r_{\max}/2\}$, i.e.
\[ \supp \psi^\text{B}(t, .) \subset [0, r_{\max}/2) \times \partial \scrN \qquad \text{for all~$t \in [0, \varepsilon)$} \]
(see the bottom picture in Figure~\ref{fig}). Therefore, extending~$\psi^\text{B}$ by zero, we obtain a global solution in all~$\scrM$.

The function~$\psi = \psi^\text{B} + \psi^\text{I}$ is the desired solution of our mixed
initial/boundary value problem. Uniqueness follows immediately from standard energy estimates
for symmetric hyperbolic systems (see for example~\cite[Section~5.3]{john}).
\QED

\begin{Corollary} \label{corfinal} The mixed initial/boundary value problem~\eqref{initboundary},
\eqref{initclass} has a unique global solution~$\psi$ in the class of smooth wave functions
with spatially compact support satisfying the boundary conditions,
\[ \Big\{ \psi \in C^\infty_\textnormal{sc}(\scrM, S\scrM) \quad \text{with} \quad (\slashed{n} - i)\:
\big( H^p \psi \big) \big|_{\partial \scrM} = 0 \quad \text{for all~$p \in \N_0$} \Big\} \:. \]
The resulting time evolution operator is unitary with respect to the scalar product
\beq \label{pint}
(\psi | \phi)_\scrN = \int_\scrN \Sl \psi(t,x) \,|\, \slashed{\nu}(t,x)\: \phi(t,x) \Sr_x \: d\mu_\scrN(x)\:.
\eeq
\end{Corollary}
\Proof Since the existence time~$\varepsilon$ in Lemma~\ref{lemmalocal} does not depend on
the initial data, we can iterate the procedure to obtain smooth solutions for arbitrarily large times.
Moreover, solving backwards in time, one can also obtain smooth solutions for arbitrarily large negative times.
We thus obtain global smooth solutions~$\psi \in C^\infty_\textnormal{sc}(\scrM, S\scrM)$.
The symmetry of~$H$ (as shown after~\eqref{domright}) implies that the scalar product~\eqref{pint}
is preserved under time evolution. Therefore, the time evolution operator is unitary.
\QED

\section{Self-Adjointness of the Dirac Hamiltonian}
We now give the proof of Theorem~\ref{thmmain}.
Let~$H$ be the Dirac Hamiltonian with domain~$\D(H)$ given by~\eqref{domright2}.
Corollary~\ref{corfinal} shows that the time evolution operator for the
mixed initial/boundary value problem~\eqref{initboundary}, \eqref{initclass}
defines a one-parameter group acting on~$\D(H)$.
Moreover, it is obvious that the domain is invariant under the action of~$H$.
Therefore, we can apply the result by Chernoff~\cite[Lemma~2.1]{chernoff73} to conclude that~$H$ is essentially
self-adjoint on~$\D(H)$. This completes the proof of Theorem~\ref{thmmain}. \\

\Thanks {{\em{Acknowledgments:}}
We would like to thank the referee for helpful comments.
F.F.\ is grateful to the Center of Mathematical Sciences and Applications at
Harvard University for hospitality and support. 

\providecommand{\bysame}{\leavevmode\hbox to3em{\hrulefill}\thinspace}
\providecommand{\MR}{\relax\ifhmode\unskip\space\fi MR }
\providecommand{\MRhref}[2]{%
  \href{http://www.ams.org/mathscinet-getitem?mr=#1}{#2}
}
\providecommand{\href}[2]{#2}

\end{document}